\global\let\ifmypprint\iffalse 
\def\mypprint{\global\let\ifmypprint\iftrue}
\global\let\iftorefs\iffalse
\def\torefs{\global\let\iftorefs\iftrue}
\global\let\dofloatfig\iffalse
\def\floatthefig{\let\dofloatfig\iftrue}
\mypprint 
\ifmypprint
\documentstyle[twocolumn,prl,aps]{revtex}
\floatthefig
\else
  \documentstyle[preprint,aps]{revtex}
  \iftorefs 
    \catcode`\@=11 
    
    \def\figure{\let\@capwidth\columnwidth\@float{figure}}
    \let\endfigure\end@float
    \@namedef{figure*}{\let\@capwidth\textwidth\@dblfloat{figure}}
    \@namedef{endfigure*}{\end@dblfloat}
    \catcode`\@=12 
     \floatthefig 
  \fi
\fi

\input epsf.tex
\begin{document}
\draft
\title{Long-Wavelength Instability in Surface-Tension-Driven 
B\'{e}nard Convection}
\author{Stephen J. VanHook\cite{email1}, Michael 
F. Schatz\cite{email2}, William D. McCormick,\\
J. B. Swift, and Harry L. Swinney\cite{email3}}
\address{Center for Nonlinear Dynamics and Department of Physics\\
University of Texas at Austin, Austin, Texas 78712}
\date{\today}
\maketitle
\widetext
\begin{abstract}
Laboratory studies reveal a deformational instability that
leads to a drained region (dry spot) in an initially flat liquid layer 
(with a free upper surface) heated 
uniformly from below.
This long-wavelength instability supplants 
hexagonal convection cells as the primary instability 
in viscous liquid layers that are sufficiently thin or are in 
microgravity.  The instability 
occurs at a temperature gradient 34\% smaller than
predicted by linear stability theory.  Numerical simulations
show a drained region qualitatively similar to that seen in the 
experiment. 
\end{abstract}
\narrowtext


B\'{e}nard's observation in 1900 \cite{Benard} of hexagonal
convection patterns launched the modern study
of convection, pattern formation, and instabilities;  
yet understanding of the surface-tension-driven
regime in which B\'{e}nard performed his experiments is still
far from complete.  
Block \cite{Block} and Pearson \cite{Pearson} 
first showed how temperature-induced 
surface tension gradients (thermocapillarity)
caused the instability observed in B\'{e}nard's experiments.  
Pearson's linear stability analysis with a 
nondeformable liquid-gas interface yielded an instability
at wavenumber $q$ = 1.99 (scaled by the mean liquid depth $d$)
and Marangoni number $M_{c} = 80$, where
$M \equiv \sigma_{T}\bigtriangleup{T}d/\rho\nu\kappa$
(see Fig. \ref{sketch}) expresses the competition between 
the destabilization by thermocapillarity and the stabilization 
by diffusion.

A deformable free surface allows a second type 
of primary instability in which a perturbation creates 
a nonuniform liquid depth and temperature profile
\cite{ScrandSter,Smith}.
Thermocapillarity causes cool, elevated regions to pull liquid from  
warm, depressed regions (see Fig. \ref{sketch}). 
The instability appears with a long wavelength since surface tension 
stabilizes short wavelengths.
Linear stability analyses that allow for deformation 
reveal this instability (see Fig. \ref{linstab}) at
zero wavenumber ($q = 0$) and $M_{c} = {2 \over 3}G$ 
\cite{Smith,Linstabs}, where the Galileo number, 
$G \equiv gd^{3}/\nu\kappa$ ($g$ is the acceleration of
gravity), gives the relative strengths of the stabilizing mechanisms of
diffusion and gravity and thus determines which
instability will form.  
For sufficiently thin, viscous liquid layers or
small $g$ ({\it e.g.}, microgravity), ${2 \over 3}G < 80$, so
the long-wavelength mode should become the primary instability 
(see Fig. \ref{linstab}).
This long-wavelength instability has not 
been experimentally investigated.  


In this Letter we describe 
experimental observations of the onset of 
the long-wavelength instability and compare these observations to
linear stability theory.  The instability leads to a large-scale
drained region with diameter $\sim$ 100$d$.  
The qualitative features of the instability are compared to nonlinear
theory based on a long-
\ifmypprint\pagebreak \vspace*{1.31in} \noindent\fi
wavelength evolution equation.  
We also explore the competition between the long-wavelength and
hexagonal instabilities and study the physical mechanism
that selects which pattern will appear.  For a range of liquid depths,
both patterns coexist.

\smallskip
\dofloatfig
\begin{figure}
\epsfxsize=8.5truecm
\centerline{\epsffile{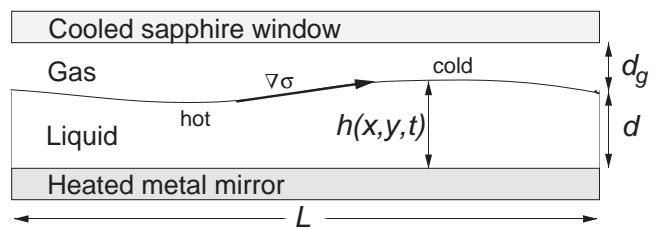}}
\smallskip
\caption{A surface-tension-driven B\'{e}nard (Marangoni) 
convection cell (not to scale) contains both a liquid (silicone oil)
layer of mean depth $d$ and local depth $h(x,y,t)$ and 
a gas (air) layer of mean thickness $d_{g}$.  
The mean temperature drop across the liquid layer is ${\Delta}T$.
The liquid has density $\rho$ (0.94 g/cm$^{3}$), 
kinematic viscosity $\nu$, 
thermal diffusivity $\kappa$ (0.0010 cm$^2$/s), and temperature
coefficient of surface tension 
$\sigma_{T} \equiv {\mid{d}\sigma{/dT}\mid}$
(0.069 dynes/cm$^\circ$C).}
\label{sketch}
\end{figure}
\fi

\dofloatfig
\begin{figure}
\epsfxsize=8.5truecm
\centerline{\epsffile{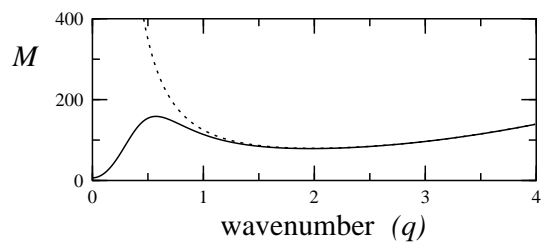}}
\smallskip
\caption{Marginal stability curves for layers of 10 cS silicone oil at two
thicknesses:  $d$ = 0.1 cm (- - -),
hexagonal convection cells form at $M_{c} = 80$ with $q$ = 1.99 
($G = 10^{4}$); $d$ = 0.01 cm (-----), the long-wavelength
($q = 0$) instability forms at $M_{c} = 6.7$ ($G = 10$).} 
\label{linstab}
\end{figure}
\fi


We study a thin layer of silicone oil that lies  
on a heated, gold-plated aluminum mirror and is 
bounded above by an air layer (see Fig. \ref{sketch}).  
A single-crystal sapphire window (0.3-cm-thick) 
above the air is cooled by 
a temperature-controlled chloroform bath.  
The temperature drop across
the liquid layer is calculated assuming conductive 
heat transport \cite{Koschmieder} 
and is typically 0.5--5 $^{\circ}$C.
We use a polydimethylsiloxane silicone oil \cite{viscosity}
with a viscosity of 10.2 cS at 50 $^\circ$C.  
The circular cell (3.81 cm inner diameter) has aluminum sidewalls
whose upper surface is made non-wetting with a coating of Scotchgard.
The experiments are performed with 0.005 cm $< d <$ 0.025 cm and
0.023 $< d_{g} <$ 0.080 cm (typically $d_{g}$ = 0.035 cm); 
the corresponding
fundamental wavevectors are in the range 
0.008 $<$ $q$ = 2${\pi}d/L$ $<$ 0.040 ($\ll$ 1).  
The liquid layers are sufficiently thin that buoyancy is negligible \cite{psi}.
The gap between the sapphire 
window and the mirror is uniform to 1\%, as verified interferometrically.
The liquid surface is initially flat and parallel to the mirror to 
1\% in the central 90\% of the cell, with a boundary region near 
the sidewalls due to contact line pinning.  
This initial depth variation is accentuated 
by thermocapillarity as $\Delta{T}$ is increased; measurements of depth
variation show a 10\% surface deformation (in the central 90\% of the cell)
at 3\% below onset.
For visualization, we use interferometry, 
shadowgraph and infrared imaging  
(256 $\times$ 256 InSb staring array, sensitive 
in the range 3-5 $\mu$m).

\smallskip
\dofloatfig
\begin{figure}[ht]
\epsfxsize=8.5truecm
\centerline{\epsffile{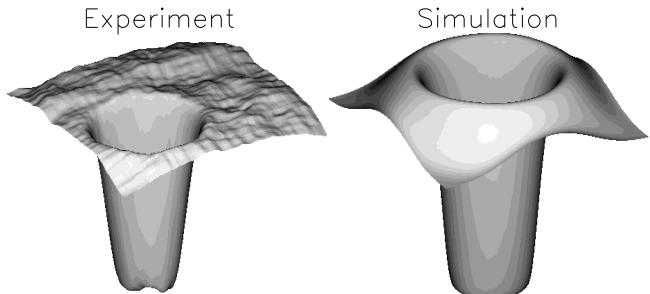}}
\smallskip
\caption{Dry spot in both experiment and
numerical simulation. The experimental picture shows
the measured brightness temperature as a function of
position along the interface just above ($\sim$ 10\%) onset; 
$d$ = 0.011 cm.  
Numerical simulation of a long-wavelength evolution equation  
shows the depth as
a function of position for 1\% above onset of linear instability. 
The depth of the liquid goes to zero in the drained region.}
\label{expsim}
\end{figure}
\fi


Above a critical $\Delta{T}$, the liquid layer becomes unstable to 
a long-wavelength draining 
mode that eventually forms a dry spot [see Fig. 
\ref{expsim}(a)]. 
The drained region takes several hours (of order a horizontal
diffusion time, $L^{2}/\kappa$) to form.
The dry spot is not completely dry since an 
adsorbed layer $\sim$ 1  $\mu$m remains \cite{Burelbach}.  The
size of the drained region is typically one quarter to one third 
the diameter of the entire cell.

Our measurements for the onset of instability are compared 
to the prediction of 
linear stability theory \cite{Linstabs,Bi1,discrete}
in Fig. \ref{theorycomp}.  The results are given in
terms of the dynamic Bond number, 
$B \equiv G/M = {\rho}gd^{2}/\sigma_{T}{\Delta}T$, 
which is the relevant control parameter for the 
long-wavelength instability;
$B$ is a measure of the balance between gravity's 
stabilizing influence and
thermocapillarity's destabilizing effect.  
Figure \ref{theorycomp} shows that the measured 
critical values of $B^{-1}$
are independent of liquid depth and viscosity, as predicted by theory.  
However, $B_{c}^{-1}$ in the experiment is 34\% smaller than
predicted. We do not believe this discrepancy is due to systematic errors
in the characterization of our experiment ({\it e.g.},
geometry, fluid properties) 
since experiments in the same convection cell 
using thicker liquid layers 
find onset of hexagons in agreement with 
another experiment \cite{Schatz1995a} and linear stability theory 
\cite{Pearson}.  

\smallskip
\dofloatfig
\begin{figure}[ht]
\epsfxsize=8.5truecm
\centerline{\epsffile{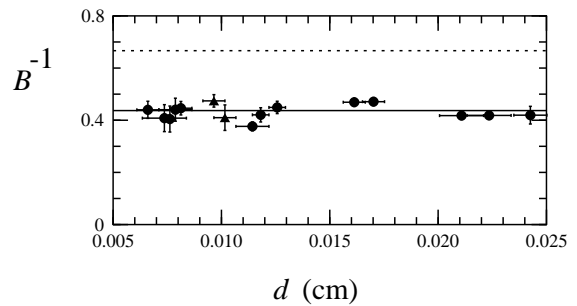}}
\smallskip
\caption{Measurements of onset compared to the 
theoretical prediction, $B_{c}^{-1} = {2 \over 3}$ (- - -).
 (Solid circles, 10.2 cS silicone oil; solid triangles, 
30 cS silicone oil.) 
The weighted mean of the data is 
$B_{c}^{-1} =$ 0.44 $\pm$ 0.03 (-----).  
There is an additional uncertainty
of $\pm$0.06 in $B_{c}^{-1}$ from the uncertainty 
in the thermal properties of the silicone oil.}
\label{theorycomp}
\end{figure}
\fi


The long-wavelength and the hexagonal modes become simultaneously 
unstable at a critical liquid depth $d_{c}$.  Near this critical depth, both
modes of instability compete and influence the formation of
the pattern.  The two modes
are not mutually exclusive, but there is a fundamental imbalance in their
relationship:   the presence of the hexagons 
suppresses the long-wavelength mode, while the 
presence of long-wavelength
deformation may induce the formation of hexagons.  
Linear theory predicts 
$d_{c} = (120\nu\kappa/g)^{1/3}$ = 0.023 $\pm$ 0.001 cm; 
we observe the exchange of primary instabilities at 
$d_{c}$ = 0.025 $\pm$ 0.001 cm.  For $d > d_{c}$
this depth, hexagons are the primary instability 
[see Fig. \ref{lwhexcom}(d)]; 
the hexagons smooth the large-scale temperature 
variations that would allow
formation of the long-wavelength mode as 
a secondary instability when $B^{-1}$ exceeds 
$B_{c}^{-1}$.  For $d < d_{c}$, 
the long-wavelength mode is the primary instability 
[see Fig. \ref{lwhexcom}(b)].  The fluid expelled from 
the forming dry spot increases the local height $h(x,y,t)$ and thus the 
local Marangoni 
number [$\propto h(x,y,t)$] in the newly 
formed elevated region; for 0.017 cm $< d < $ $d_c$,
hexagons form in the elevated region since the local $M$ in this region 
exceeds $M_{c}$ (80) for the onset of hexagons [see Fig. 
\ref{lwhexcom}(c)].  For $d<$ 0.017 cm, hexagons 
do not form at the onset of the 
long-wavelength mode, but can form in the elevated region for 
$\Delta{T}$ sufficiently above onset.  
Similar mode competition phenomena have been studied theoretically for 
solutocapillary convection \cite{GNP1}.
When $d \ll d_{c}$, increasing $\Delta{T}$ above $\Delta{T_{c}}$
increases the area of the dry spot; no qualitatively
new structures are observed.  At fixed $\Delta{T}$, the dry spot
is stable.

\smallskip
\dofloatfig
\begin{figure}[ht]
\epsfxsize=8.5truecm
\centerline{\epsffile{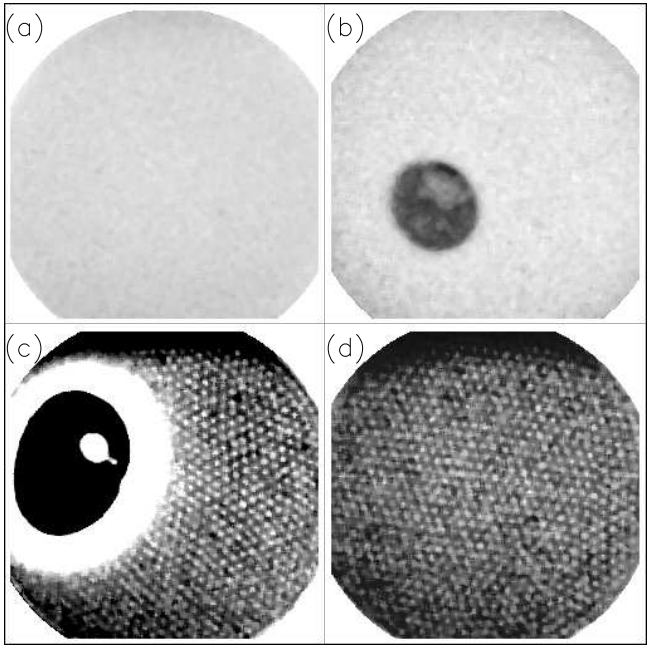}}
\smallskip
\caption{Infrared images from experiments
with increasing fluid depth (increasing $G$).  
(a) 10\% below onset of instability ($d$ = 0.033 cm). 
(b) For $d$ = 0.011 cm, the long-wavelength mode is the primary instability.
The dark region is the dry spot.  (c) At $d$ = 0.022 cm, the
long-wavelength and hexagonal modes coexist.  A droplet (white circle)
is trapped within the dry spot (dark oval); the liquid layer is strongly 
deformed (white annulus) between the dry spot and the hexagonal pattern.
 (d) For $d$ $>$ 0.025 cm, hexagons are the primary instability 
(here, $d$ = 0.033 cm).}
\label{lwhexcom}
\end{figure}
\fi


To understand the long-wavelength instability more 
fully, we study a long-wavelength 
evolution equation derived by 
Davis\cite{Davis83,Davis87} for an insulating
upper boundary and later considered by other authors 
\cite{KandP,OandR} for more general upper boundary conditions.  
The evolution 
equation for the local liquid depth $h(x,y,t)$ (scaled by $d$) 
with an insulating upper boundary is
$${3 \over G}h_t + \bigtriangledown\cdot\left\{
{3 \over 2{B}}h^{2}{\bigtriangledown}h
-h^3\bigtriangledown h + 
{(2{\pi})^2 \over B_{o}}
h^3\bigtriangledown^{2}\bigtriangledown{h}
\right\} = 0$$
where the domain of both $x$ and $y$ is [0,2$\pi$), 
time is scaled by $d^2/\kappa$,
$B$ is the dynamic Bond number as before, 
and the static Bond number is $B_{o} \equiv {\rho}gL^{2}/{\sigma}$.
The first term in curly brackets describes the effect of 
thermocapillarity; the second, 
gravity; and the third, surface tension.  
A linear stability analysis of this equation agrees with the linear stability
analysis of the full fluid equations referenced earlier.  
Numerically integrating the evolution equation in time 
using a pseudo-spectral code
with periodic boundary conditions (which automatically satisfy the 
incompressibility 
condition), we find that when the system is unstable, an initially 
infinitesimal perturbation grows until it forms a dry spot, as shown
in Fig. \ref{expsim}(b) ($B_{o}$ = 1200 in the simulation, 
700 in the experiment).  Dry spots in the simulation
do not saturate (unlike the observed spots),  
since the assumptions leading to the evolution equation are violated 
once the dry spot has formed; in this regime, the simulation 
quickly loses spectral convergence and breaks down.  
The shape of the dry spot near onset is independent of parameter values
except for $B_{o}$, which governs the size of the drained 
region.  For large $B_{o}$, sharp 
structures can form; for small $B_{o}$, rounding by surface tension
prevents the formation of sharp structures.  


We find no stable, intermediate solution of the evolution equation between
no deformation and a drained region.  
Our weakly nonlinear analysis of the evolution equation predicts a
subcritical instability for all parameter values.  
We have performed one- and two-dimensional simulations as well as an
analysis of 
the evolution equation to look for any turnover in the backwards unstable 
(subcritical) branch of the bifurcation.
For an insulating upper boundary, we have shown 
analytically that no stable deformed 
solutions exist, in agreement with previous authors \cite{OandR}.  
The unstable branch continues backwards until 
$B^{-1} = {1 \over 3}$, at which point it 
disappears since the unstable solution past that point is unphysical ($h < 0$); 
nowhere does the backwards unstable curve turn over in a saddle-node 
bifurcation.  Simulations using more 
general upper thermal boundary conditions \cite{Bi1} 
also find no stable deformed solutions.


In conclusion, we have observed the formation of a large-scale
dry spot due to a long-wavelength deformational instability.  
The formation of such dry spots could be a serious problem in the
planned use of liquid layers (even as thick as one centimeter) in 
microgravity environments, where fluid motion is driven primarily
by surface tension gradients (buoyancy effects are negligible).  

The structure of the dry spot agrees qualitatively with numerical
simulations.  Linear stability theory correctly predicts the functional
dependence of the onset on experimental parameters, but 
the onset in the experiment occurs at 
$B_{c}^{-1}$ (or ${\Delta}T_{c}$) 34\% 
smaller than predicted by linear theory.
This discrepancy in the onset may be due to the difference 
in lateral boundaries
between theory and experiment.  
For example, all  theory to date has assumed
periodic boundary conditions, which become suspect when the wavelength
of the mode under consideration is of order the system size.
The long-wavelength
evolution equation is not valid near a boundary and it is not clear
what boundary conditions other than periodic might 
be used for the evolution
equation.  Future numerical simulations employing realistic, 
finite horizontal boundaries may answer this question.

We thank S. H. Davis, R. E. Kelly, and E. L. Koschmieder 
for useful discussions.  This research is supported by
the NASA Microgravity Science and Applications 
Division (Grant No. NAG3-1382).  
S.J.V.H. is supported by the 
NASA Graduate Student Researchers Program.


%
%
\dofloatfig\else
\begin{figure}
\smallskip
\caption{A surface-tension-driven B\'{e}nard (Marangoni) 
convection cell (not to scale) contains both a liquid (silicone oil)
layer of mean depth $d$ and local depth $h(x,y,t)$ and 
a gas (air) layer of mean thickness $d_{g}$.  
The mean temperature drop across the liquid layer is ${\Delta}T$.
The liquid has density $\rho$ (0.94 g/cm$^{3}$), 
kinematic viscosity $\nu$, 
thermal diffusivity $\kappa$ (0.0010 cm$^2$/s), and temperature
coefficient of surface tension 
$\sigma_{T} \equiv {\mid{d}\sigma{/dT}\mid}$
(0.069 dynes/cm$^\circ$C).}
\label{sketch}
\end{figure}

\begin{figure}
\smallskip
\caption{Marginal stability curves for layers of 10 cS silicone oil at two
thicknesses:  $d$ = 0.1 cm (- - -),
hexagonal convection cells form at $M_{c} = 80$ with $q$ = 1.99 
($G = 10^{4}$); $d$ = 0.01 cm (-----), the long-wavelength
($q = 0$) instability forms at $M_{c} = 6.7$ ($G = 10$).} 
\label{linstab}
\end{figure}

\begin{figure}[ht]
\smallskip
\caption{Dry spot in both experiment and
numerical simulation. The experimental picture shows
the measured brightness temperature as a function of
position along the interface just above ($\sim$ 10\%) onset; 
$d$ = 0.011 cm.  
Numerical simulation of a long-wavelength evolution equation  
shows the depth as
a function of position for 1\% above onset of linear instability. 
The depth of the liquid goes to zero in the drained region.}
\label{expsim}
\end{figure}

\begin{figure}[ht]
\smallskip
\caption{Measurements of onset compared to the 
theoretical prediction, $B_{c}^{-1} = {2 \over 3}$ (- - -).
 (Solid circles, 10.2 cS silicone oil; 
solid triangles, 50 cS silicone oil.]  
The weighted mean of the data is 
$B_{c}^{-1} =$ 0.44 $\pm$ 0.03 (-----).  
There is an additional uncertainty
of $\pm$0.06 in $B_{c}^{-1}$ from the uncertainty 
in the thermal properties of the silicone oil.}
\label{theorycomp}
\end{figure}

\begin{figure}[ht]
\smallskip
\caption{Infrared images from experiments
with increasing fluid depth (increasing $G$).  
(a) 10\% below onset of instability ($d$ = 0.033 cm). 
(b) For $d$ = 0.011 cm, the long-wavelength mode is the primary instability.
The dark region is the dry spot.  (c) At $d$ = 0.022 cm, the
long-wavelength and hexagonal modes coexist.  A droplet (white circle)
is trapped within the dry spot (dark oval); the liquid layer is strongly 
deformed (white annulus) between the dry spot and the hexagonal pattern.
 (d) For $d$ $>$ 0.025 cm, hexagons are the primary instability 
(here, $d$ = 0.033 cm).}
\label{lwhexcom}
\end{figure}
\fi
\ifmypprint\else
 \vfill\eject
 \epsfxsize=6 truein  \centerline{\epsffile{vanhook1.eps}}
  \vfill{\small VanHook {\it et al.}, Fig.~1} \vfill\eject
 \epsfysize=6.0truein  \centerline{\epsffile{vanhook2.eps}}
 \vfill{\small VanHook {\it et al.}, Fig.~2} \vfill\eject \epsfysize=6truein
 \centerline{\epsffile{vanhook3.eps}}
 \vfill{\small VanHook {\it et al.}, Fig.~3} \vfill\eject
 \epsfxsize=6truein \centerline{\epsffile{vanhook4.eps}}
 \vfill{\small VanHook {\it et al.}, Fig.~4} \vfill\eject
 \epsfxsize=6truein \centerline{\epsffile{vanhook5.eps}}
 \vfill{\small VanHook {\it et al.}, Fig.~5} \vfill\eject
\fi

%
%


\begin{references}
\bibitem{discrete}
These results are for an infinite system.  
A finite system has a lowest order mode of
$q$ = 2${\pi}d/L$, which becomes unstable at a slightly higher (6\%) 
value of the control parameter.  
We do not include this correction in Fig. \ref{theorycomp}.
\end{references}
\end{document}